\documentclass[draftcls, onecolumn, 12pt, twoside]{IEEEtran}

\IEEEoverridecommandlockouts

\usepackage{algorithm}
\usepackage{algorithmic}
\usepackage{amsfonts}
\usepackage{amsthm}
\usepackage{amssymb}
\usepackage{array}
\usepackage{bbm}
\usepackage{bbold}
\usepackage{bm}
\usepackage{booktabs}
\usepackage{caption}
\usepackage{cite}
\usepackage[cmex10]{amsmath}
\usepackage{diagbox}
\usepackage{enumerate}
\usepackage{epstopdf}
\usepackage{footnote}
\usepackage{framed}
\usepackage{graphicx}
\usepackage{indentfirst}
\usepackage{lineno}
\usepackage{multicol}
\usepackage{multirow}
\usepackage{setspace}
\usepackage{subeqnarray}
\usepackage{subfigure}
\usepackage{threeparttable}
\makesavenoteenv{tabular}
\usepackage{url}
\usepackage{xcolor}
\newcommand{\mb}{\mathbf}
\newcommand{\mbb}{\mathbb}
\newcommand{\mc}{\mathcal}

\newcommand{\ti}{\textit}

\newtheorem{theorem}{\textbf{Theorem}}
\newtheorem{lemma}{\textbf{Lemma}}

\newtheorem{definition}{\textbf{Definition}}

\newtheorem{construction}{\textbf{Construction}}

\begin{document}
\begin{sloppypar}

\title{Two Classes of Optimal
Multi-Input Structures for Node Computations in Message Passing Algorithms}

\author{%
\IEEEauthorblockN{Teng Lu, Xuan He, and Xiaohu Tang}\\
\IEEEauthorblockA{Information Coding and Transmission Key Lab of Sichuan Province, \\Southwest Jiaotong University, Chengdu {\rm 611756}, China\\
Email:  luteng@my.swjtu.edu.cn, xhe@swjtu.edu.cn, xhutang@swjtu.edu.cn}
}

\maketitle

\begin{abstract}
In this paper, we delve into the computations performed at a node within a message-passing algorithm.
We investigate low complexity/latency multi-input structures that can be adopted by the node for computing outgoing messages $\mb{y} = (y_1, y_2, \ldots, y_n)$ from incoming messages $\mb{x}  = (x_1, x_2, \ldots, x_n)$, where each $y_j, j = 1, 2, \ldots, n$ is computed via a multi-way tree with leaves $\mb{x}$ excluding $x_j$.
Specifically, we propose two classes of structures for different scenarios.
For the scenario where complexity has a higher priority than latency, the star-tree-based structures are proposed.
The complexity-optimal ones (as well as their lowest latency) of such structures  are obtained, which have the near-lowest (and sometimes the lowest) complexity among all structures. 
For the scenario where latency has a higher priority than complexity, the isomorphic-directed-rooted-tree-based structures are proposed.
The latency-optimal ones (as well as their lowest complexity) of such structures  are obtained, which are proved to have the lowest latency among all structures.

\end{abstract}


\IEEEpeerreviewmaketitle

%%%%%%%%%%%%%%%%%%%%%%%%%%%%%%%%%%%%%%%%%%%%%%%%%%%%%%%%%%%%%%%%%%%%%%%%%%%%%%%%%%%%%%%%%%%%%%%%%%%%%%%%%%%%%%%%%%%%%%%%%%%%%%%%%%%%%

\section{Introduction}\label{section: introduction}

A large variety of algorithms in the areas of coding, signal processing, and artificial intelligence can be viewed as instances of the message passing algorithm \cite{richardson2008modern, Frank01Factor}.
  Specific instances of such algorithms include Kalman filtering and smoothing \cite{ArasaratnamH09Cubature,XuZ21FAST,YuM24Robust}; the forward-backward algorithm for hidden Markov models \cite{GauvainL94Maximum, GhahjaverestanM16Coupled, GuoLZSX21predictive}; probability propagation in Bayesian networks \cite{SunZY06bayesian, ZengZQLZ24Bayesian}; and decoding algorithms for error-correcting codes such as low-density parity-check (LDPC) codes  \cite{Gallager62, MacKay99, Richardson01capacity, 5gChannel1,  ding2023anefficient, Chen05, Ueng2017}.
These algorithms operate under a graphical framework, where each node collects incoming messages via its connecting edges, computes outgoing messages, and propagates back outgoing messages via the same edges.

\begin{figure}[h]
\centering
\includegraphics[scale = 0.5]{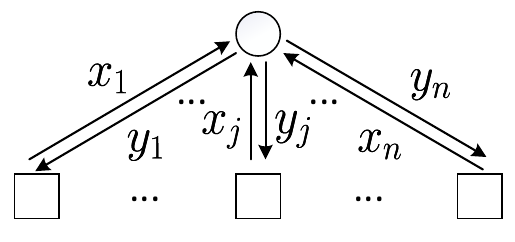}
\caption{Message passing on a specific node.}
\label{fig: node_update}
\end{figure}

In particular, we focus on the message passing on a specific node.
%For example, Fig. xx, performs the message passing in a specific node.
As shown in Fig. \ref{fig: node_update}, the incoming messages to this node are expressed as a vector $\mb{x}  = (x_1, x_2, \ldots, x_n)$,
where $x_j$ comes from the $j$-th connecting edge, for each $j \in [n] \triangleq \{1, 2, \ldots, n\}$.
Subsequently, the node computes $n$ outgoing messages, represented as $\mb{y} = (y_1, y_2, \ldots, y_n)$, where $y_j \triangleq f(x_1,...,x_{j-1},x_{j+1},...,x_n)$ is the outgoing message sent back to the $j$-th connecting edge and $f$ denotes a general function for computing outgoing messages.
Note that when computing $y_j$, the function $f$ does not include $x_j$ as input.
This is a well-known computation principle in message passing algorithms to avoid propagating $x_j$'s messages back to it again.

\begin{figure*}[t!]
\centering
\includegraphics[scale = 0.45]{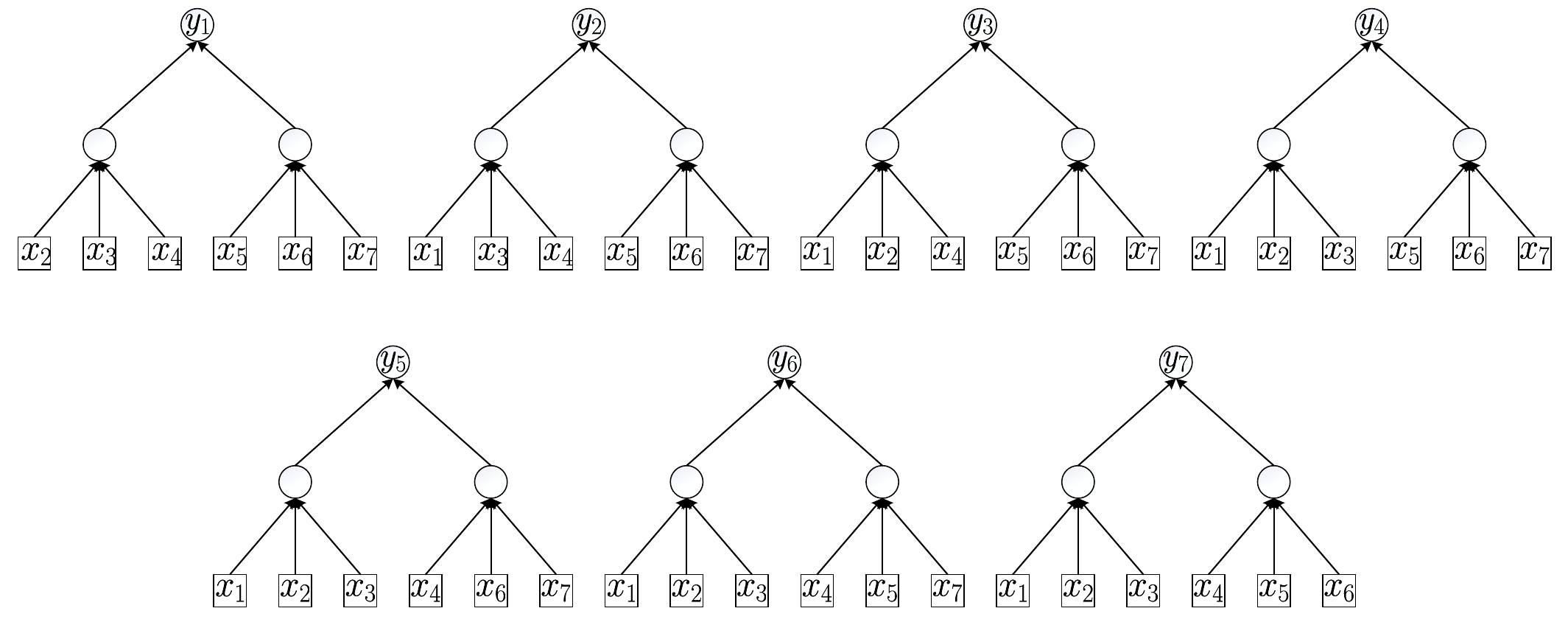}
\caption{DRTs rooted at $\mb{y}$ with $n=7$.}
\label{fig: n7_non-optimal_DRTs}
\end{figure*}

For any $j \in [n]$, the processing of computing $y_j$ can be abstracted as a directed rooted tree (DRT), whose root is $y_j$ and leaves are $\mb{x}$ excluding $x_j$.
In such a DRT, each $i$-input internal node corresponds to an $i$-input operator.
For example, assume $n=7$ and $y_j = f(x_1,...,x_{j-1},x_{j+1},...,x_n),$ $ \forall j \in [n]$.
The computation of $\mb y$ can be handled by the  seven DRTs shown in Fig. \ref{fig: n7_non-optimal_DRTs}.
The processing of computing $y_1$ can be represented as   $y_1 = f( f (x_2, x_3, x_4), f(x_5, x_6, x_7) )$, corresponding to the first DRT in Fig. \ref{fig: n7_non-optimal_DRTs}.
The computation of any other $y_j$ is analogous, and all these DRTs can be handled in parallel.
As a result, the complexity for computing $\mb{y}$ by such seven DRTs can be measured by fourteen 3-input and seven 2-input nodes, and the latency  can be measured by one 3-input and one 2-input nodes.

\begin{figure}[h]
\centering
\includegraphics[scale = 0.5]{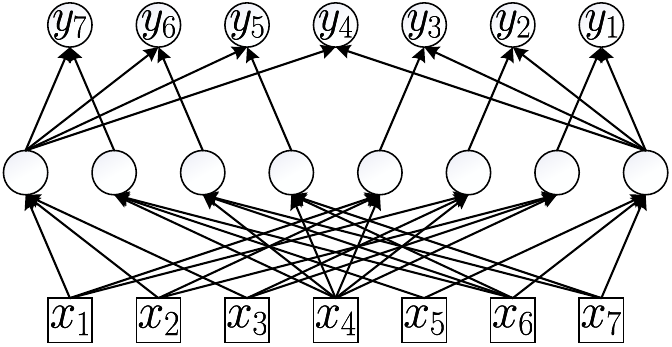}
\caption{A structure in \cite{Ueng2017} for computation of $\mb{y}$ with $n=7$.}
\label{fig: S7_structure_Ueng2017}
\end{figure}

In fact, we can reuse the computation units corresponding to the same subtree among different DRTs, so that the complexity can be reduced under the same latency.
Formally, we use a class of graphs, called structure to represent this process.
For instance, the seven DRTs in Fig. \ref{fig: n7_non-optimal_DRTs} can be united into the structure in Fig. \ref{fig: S7_structure_Ueng2017}, such that this structure realizes the same computation of $\mb y$ but requires lower complexity under the same latency.
More specifically, the complexity of the structure can be measured by  eight 3-input and seven 2-input nodes.
We remark that, the structure in Fig. \ref{fig: S7_structure_Ueng2017} was indeed proposed in \cite{Ueng2017} for efficiently implementing the node update in stochastic LDPC decoders.

\begin{figure}[t]
\centering
\includegraphics[scale = 0.5]{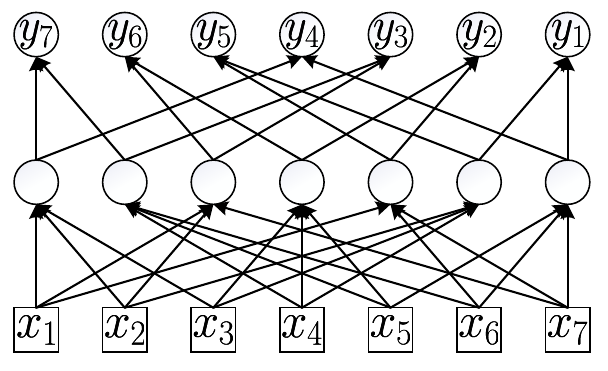}
\caption{A latency-optimal isomorphic-DRT-based structure for computation of $\mb{y}$ with $n=7$.}
\label{fig: n7_structure}
\end{figure}

However, the latency and complexity may vary in different structures.
For example, the computing process corresponding to the structure in Fig. \ref{fig: n7_structure} can also obtain $\mb y$ with $n = 7$.
The complexity of the structure can be measured by seven 3-input and seven 2-input nodes, and the latency can be measured by one 3-input and one 2-input nodes.
Therefore, under the same latency, the structure has a lower complexity (one less 3-input node) than that in Fig. \ref{fig: S7_structure_Ueng2017}.
It is always of significant interest to find structures with lower complexity/latency.
To this end, we \cite{he22aclass} investigate a class of binary-input structures, but the systematic construction of low complexity and/or low latency muti-input  structures remains open.
Here ``multi-input'' means that there may exist $k$-input computation nodes for any $k \in  [2, m] \triangleq \{2, 3, ..., m\}$, where $m \geq 2$ is the maximum number of inputs for a node and it should be regarded as a preset and fixed integer throughout this paper.

In this paper, we focus on low complexity/latency multi-input structures for computing $\mb{y}$ from $\mb{x}$.
Specifically, we propose two classes of multi-input structures for different scenarios:
\begin{itemize}

    \item For the scenario where complexity has a higher priority than latency, we propose a class of structures, called star-tree-based structures.
Among these structures, we derive algorithms for computing their lowest complexity as well as  the lowest latency of the complexity-optimal structures. We derive an algorithm to obtain the complexity-optimal ones of such structures, and illustrate that they have the   near-lowest (and sometimes the lowest) complexity among all structures.  We further derive an algorithm to obtain the latency-optimal ones among the complexity-optimal star-tree-based structures.

\item For the scenario where latency has a higher priority than complexity, we propose a class of structures, called isomorphic-DRT-based structures.
We derive an algorithm to obtain the latency-optimal ones of such structures, and prove that they have the lowest latency among all structures. We further develop a construction for the complexity-optimal ones among the latency-optimal isomorphic-DRT-based structures.
\end{itemize}

The remainder of this paper is organized as follows.
Firstly, Section  \ref{section: preliminary} presents preliminaries related to graphs and trees.
 Next, Section \ref{section: Structures}  defines the  structures under consideration for node computation in this paper.
Then, Sections \ref{section: Star-Tree-Based Structures} and \ref{section: Isomorphic-DRT-Based Structures} delve into star-tree-based and isomorphic-DRT-based structures, respectively.
Finally, Section \ref{section: Conclusion} concludes this paper.

\section{Preliminaries}\label{section: preliminary}

In this section, we introduce preliminaries regarding graphs and trees, mainly based on \cite{cormen2009introduction}.
For  convenience, define $\mathbb{Z}_{+} \triangleq \{0, 1, \ldots\}$ as the set of non-negative integers.
For any $i, j \in \mathbb{Z}_{+}$ with $i < j$, define $[i, j] \triangleq \{i, i+1, \ldots, j\}$ and $[j] \triangleq \{1, 2, \ldots, j\}$.
For any $i \in [m-1]$, let $\mb e_i$ be a vector of length $m-1$ whose $i$-th entry is 1 and other entries are 0.

\subsection{Graphs}

A graph, denoted as $G\triangleq G_v \cup G_e$, consists of a node set $G_v$ and an edge set $G_e$.
Any element in $G_e$ is called a directed
(resp. undirected) edge which is denoted by an ordered (resp.
unordered) pair $(a, b) \in G_v \times G_v$. The term ``ordered'' (resp.
``unordered'') implies that $(a, b) \neq (b, a)$ (resp. $(a, b) = (b, a)$).
A graph $G$ is a directed (undirected) graph if and only if (iff) each element in $G_e$ is a directed (undirected) edge.
When graphically represented, directed edges are depicted using arrows, whereas undirected edges are depicted using lines.
A graph $G'$ is called a subgraph of $G$ if $G' \subseteq G$.

In a directed graph $G$, an edge $(a, b)$ is referred to as a leaving or outgoing edge of node $a$, and an entering or incoming edge of node $b$.
In contrast, for an undirected graph $G$, the edge $(a, b)$ is simply described as an edge of $a$ and $b$.
To describe the operations of removing or adding a node $a$ or an edge $(a, b) $ from/to a graph $G$, we utilize the notation of subtraction/addition (represented by $-/+$).
Specifically, $G - a \triangleq G \setminus( \{a\} \cup \{(a_1, a_2) \in G: a_1 = a ~\text{or}~ a_2 = a\})$,
$G + a \triangleq G \cup \{a\}$,
$G - (a, b) \triangleq G \setminus \{(a, b)\}$,
and $G + (a, b) \triangleq G\cup\{a, b\} \cup \{(a, b)\}$.

For any graph $G$ and node $a \in G$, define $d(a,G)$ as the degree of $a$ in $G$.
If $G$ is an undirected graph, $d(a,G)$ is the number of edges connecting with $a$.
If $G$ is a directed graph, $d(a,G)$ is the number of incoming edges of $a$.
For any graph $G$ and $i \in \mbb{Z}_+$, define $d_i(G)$ as the number of degree-$i$ nodes in $G$.
For any graph $G$, define $\mb q \triangleq (q_1,q_2,...,q_{m-1}) \in \mathbb{Z}_{+}^{m-1}$ as the degree vector of $G$, where for each $i \in [m-1]$, $q_i=d_{i+2}(G)$ if $G$ is undirected and $q_i=d_{i+1}(G)$ if $G$ is directed.

We call node sequence $P = (v_0, v_1, \ldots, v_k)$ a path in $G$ iff $(v_{i-1}, v_i) \in G, \forall i \in [k]$, where $k$ is called the length of $P$.
Moreover, $P$ is called a simple path iff $v_i \neq v_j, \forall~ 0 \leq i < j \leq k$.
For any $a, b \in G$, we say that $a$ is reachable from $b$ if there exists a path from $b$ to $a$.
The distance from $a$ to $b$  is defined as the length of the shortest path from $a$ to $b$.
If no such path exists, the distance is regarded as $\infty$.
We further call $P$  a cycle if $k \geq 2$, $v_0 = v_k$, $(v_0, v_1) \neq (v_1, v_2)$, and $(v_1, v_2, \ldots, v_k)$ is a simple path.

For any directed graph $G$, we call $G$ a directed acyclic graph (DAG) iff there exists no cycle in $G$.
For any DAG $G$ and $a \in G$, define $E(a,G) \triangleq (\{b \in G: \text{$a$ is reachable from $b$}\}, \{(b', b'') \in G: \text{$a$ is reachable from $b''$} \})$ as a subgraph of $G$.
Moreover, we define $E(G) \triangleq \{E(a, G): a \in G\}$.
An undirected graph is connected if every node is reachable from all other nodes.

\subsection{Trees}

A tree is defined as a connected, acyclic, undirected graph.
Given any tree $T$ and any node $a \in T$, we call $a$  a leaf of $T$ if its degree $d(a,T)=1$.
Otherwise, we call $a$  an internal node.

A rooted tree is a tree in which there is a unique node called
the root of the tree.
 We consider a rooted tree $T$, with its root denoted by $r(T)$.
 The depth of any node $a \in T$ is defined as the distance between $a$ and $ r(T)$.
Additionally, for a non-negative integer $i$, denote the $i$-th level as the set of all  the nodes with the depth $i$.
The height of $T$ is determined by the greatest depth among all its nodes.
For any edge $(a,b) \in T$, if the depth of $a$ is greater than the depth of $b$, then $b$ is designated as the parent of $a$, and conversely, $a$ is the child of $b$.

The directed variant of a rooted tree $T$, say $T'$, is created by converting each undirected edge $(a,b) \in T$, where $a$ is a child of $b$, into a directed edge from $a$ to $b$ in $T'$.
 We refer to $T'$ as a directed rooted tree (DRT), with $T$ being its undirected counterpart.
For any node $a \in T'$, $E(a, T')$ is called the subtree of $T'$ rooted at $a$.
For any $(a, b) \in T$, let $D(a, b, T) \triangleq E(a, T_b)$, where $T_b$ is the result (a DRT) by making $T$ as a DRT with root $b$.
Let $D(T) \triangleq \{D(a,b, T): (a,b) \in T\}$.

\subsection{Labels of Graphs}

For any graph $G$, we call $G$ a labelled graph iff every node in $G$ is given a unique label (as a result, each edge is also given a unique label).
Conversely, if not all nodes in $G$ are uniquely labelled, $G$ is termed a partially unlabelled graph.

Two labelled graphs $G$ and $G'$ are considered identical, i.e., $G = G'$, iff $G$ and $G'$ have same labelled nodes and edges (and root for rooted trees).
Meanwhile, two partially unlabelled graphs, $G$ and $G'$, are deemed identical if there exists a labelling scheme for the unlabelled nodes in both $G$ and $G'$ that renders them labelled and identical.

Two partially unlabelled graphs $G$ and $G'$ are the same iff there exists a way to label all unlabelled nodes in $G$ and $G'$ such that $G$ and $G'$ become labelled and the identical.
Graphs $G$ and $G'$ are isomorphic if there is a way to reassign labels to all nodes in both graphs, resulting in identically labelled $G$ and $G'$.

\section{Structures for Node Computation}\label{section: Structures}

Recall that $\mb{x}  = (x_1, x_2, \ldots, x_n)$ and $\mb{y} = (y_1, y_2, \ldots, y_n)$ represent the incoming and outgoing messages, respectively.
%In this paper, we consider the case where for $j \in [n]$, a DRT $T_j$ is used to describe the computation of $y_j$ from $\mb{x}$ excluding $x_j$.
In this paper, we focus on the scenario where for each $j \in [n]$, a DRT $T_j$ is employed to delineate the computation of $y_j$ from $\mb{x}$ excluding $x_j$.
Specifically, within $T_j$, the leaves correspond to the incoming messages $\mb{x}$ excluding $x_j$, the internal nodes correspond to operations whose number of inputs is between 2 and $m$, and the root $r(T_j)$ corresponds to $y_j$.
Some examples of such DRTs for $n = 7$ are depicted in Fig. \ref{fig: n7_non-optimal_DRTs}.

Define an input set $X = \{x_j: j \in [n]\}$ and an output set $Y = \{y_j: j \in [n]\}$, where node $x_j$ (resp. $y_j$) corresponds to the $j$-th incoming message $x_j$ (resp. outgoing message $y_j$).
%is called the $j$-th input (resp. output) node which
In this paper, we remark that for any graph $G$ and any node $a \in G$, $a$ is labelled in $G$ iff $a$ is an input node from $X$ or an output node from $Y$.
Consequently, $G$ is partially unlabelled if $G_v \setminus (X\cup Y) \neq \emptyset$.
To describe the computation process, we introduce the structure defined as follows.

\begin{definition} \label{definition: structure}
For any DAG $S$, we say that $S$ is a structure with input size $n$ iff $S$ fulfills the following properties.
\begin{itemize}
\item [1)]  $S$ contains $n$ nodes with no incoming edges, which are exactly $X = \{x_1, x_2, \ldots, x_n\}$.
\item [2)]  $S$ contains $n$ nodes with no outgoing edges, which are exactly $Y = \{y_1, y_2, \ldots, y_n\}$.
\item [3)]
For $j \in [n]$, $E(y_j, S)$ is a DRT whose leaves are exactly $X \setminus \{{x_j}\}$.
%\item
\item [4)]  For two different nodes $a, b \in S$, $E(a, S) \neq E(b, S)$.

\item [5)] For any node $a \in S \setminus X$, we have $2 \leq d(a,S) \leq m$.

\end{itemize}
\end{definition}

We remark that Definition \ref{definition: structure} corresponds to \cite[Definition 2]{he22aclass} in a special case where $m=2$.
For any $n\geq 2$, define $\mc{S}_n$ as the set of all structures with input size $n$.
We may sometimes use subsets of a structure, referring to as substructures, which is defined below.

\begin{definition}\label{definition: substructure}
A DAG $S'$ is called a substructure iff there exists a structure $S$ satisfying that $S' \subseteq S$.
For any two substructures $S_1$ and $S_2$,  denote the union of $S_1$ and $S_2$ by $S_1 \vee S_2$, where only one copy of the same subtrees is kept, corresponding to the fourth property of Definition \ref{definition: structure}.
\end{definition}

According to Definition \ref{definition: substructure}, for any two substructures $S_1$ and $S_2$, $S_1 \vee S_2$ is still a substructure if $S_1 \cap S_2 \cap Y = \emptyset$.
Moreover, for any structure $S \in \mc{S}_n$, $E(y_j, S), \forall j \in [n]$ is a substructure of $S$, and $S = \vee_{j \in [n]} E(y_j, S)$ always holds.
A more specific example is that, the seven DRTs depicted in Fig. \ref{fig: n7_non-optimal_DRTs} are substructures whose union  (under the operation $\vee$) results in the structure in Fig. \ref{fig: S7_structure_Ueng2017}.

In practice, each degree-$i$ node corresponds to an $i$-input computation unit, the complexity and latency of which should increase as $i$ increases.
Therefore, we define the complexity factor and latency factor to describe this fact.
Specifically, for each $i \in [0, m]$, denote $c_i$  and $l_i$ respectively as the complexity factor and latency factor of the $i$-input computation node.
The complexity and latency factors are determined by the actual situation of the hardware.
In particular, $c_0,c_1,l_0$ and $l_1$ are set to 0, and it is reasonable to assume $c_i\leq c_j$ and $l_i \leq l_j$ for $0 \leq i < j \leq m$.
Accordingly, we can define the complexity and latency of a substructure (as well as structure) as follows.

\begin{definition}\label{definition: complexity latency}
For any substructure $S$, the complexity of $S$, denoted by $c(S)$, is defined as the weighted number of nodes in $S$:
$$c(S) \triangleq \sum_{i=2}^{m}d_i(S)c_i.$$
\end{definition}

\begin{definition}\label{definition: latency}
For any substructure $S$,
the latency of $S$ is defined as $$l(S) \triangleq \max_{P \subseteq S} {l(P,S)},$$
where  $P=(v_0,v_1,\ldots,v_k)$ is a path in $S$ and $l(P,S) \triangleq \sum_{i=0}^{k}l_{d(v_i,S)}$ is the latency of $P$.
\end{definition}

As an example, the complexity and latency of the structure in Fig. \ref{fig: n7_structure} are $7c_2+7c_3$ and $l_2+l_3$, respectively.
It is reasonable to use complexity and latency as two key criteria for evaluating the performance of a structure.
In this paper, our main purpose is to discover the structures with low  complexity and/or latency.

\section{Star-Tree-Based Structures} \label{section: Star-Tree-Based Structures}

In this section, we construct structures for the scenario where complexity has a higher priority than latency.
To this end, we focus on a class of structures, called star-tree-based structures, which probably have the (near) lowest complexity among all structures.
Specifically, Section \ref{subsection: IV-A} proposes a class of  trees, called star trees, and further presents a one-to-one mapping between star trees and star-tree-based structures.
Next, Section \ref{subsection: IV-B} establishes a necessary and sufficient condition between degree vectors and the existence of star trees. 
Thus, the complexity of a star-tree-based structure is completely determined by the degree vector of the corresponding star tree.
Therefore, Theorem \ref{theorem: co-algorithm} is proposed to find the degree vectors that lead to the complexity-optimal star-tree-based structures.
Finally, Section \ref{subsection: IV-C}  derives Theorem \ref{theorem: main program} to find a latency-optimal star-tree-based structure $S$ corresponding to a given degree vector $\mb{q}$.
As a result, if $\mb{q}$ is the uniquely optimal degree vector obtained from Theorem \ref{theorem: co-algorithm},  then $S$ is a complexity-optimal star-tree-based structure which also has the lowest latency among all complexity-optimal star-tree-based structures.

\subsection{Star Trees and Star-Tree-Based Structures}\label{subsection: IV-A}

In this subsection, we propose a class of trees, called star trees, in Definition \ref{def: star tree}.
Next, Construction \ref{construction: star_tree_to_structure} and Lemma \ref{lemma: h(T)_str} illustrate that, based on any star tree $T$, we can obtain a low complexity structure, denoted as $h(T)$, which is called a star-tree-based structure.

\begin{definition}[star trees]\label{def: star tree}
An $n$-input star tree $T$ is an (undirected) tree fulfilling the following properties.
\begin{itemize}
\item   $T$ has $n$ leaves, which are exactly $X$.
\item   The degree of each internal node in $T$ is between 3 and $m +1$. 
\end{itemize}
\end{definition}

\begin{figure}[h]
\centering
\includegraphics[scale = 0.5]{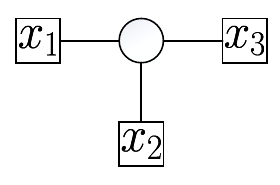}
\caption{The only star tree $T \in \mc T_3$.}
\label{fig: degree3_n7_star_tree}
\end{figure}

\begin{figure}[h]
\centering
\includegraphics[scale = 0.5]{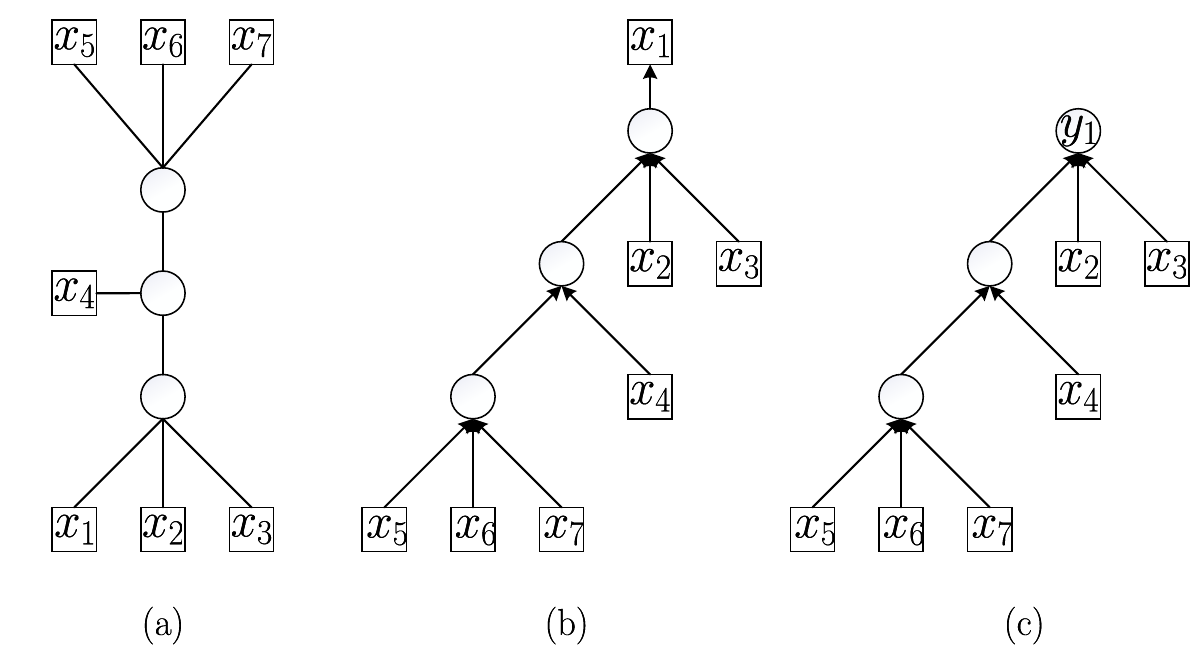}
\caption{An example for the transformation from star tree to DRTs.}
\label{fig: n7_star_tree}
\end{figure}

\begin{figure*}[t!]
\centering
\includegraphics[scale = 0.5]{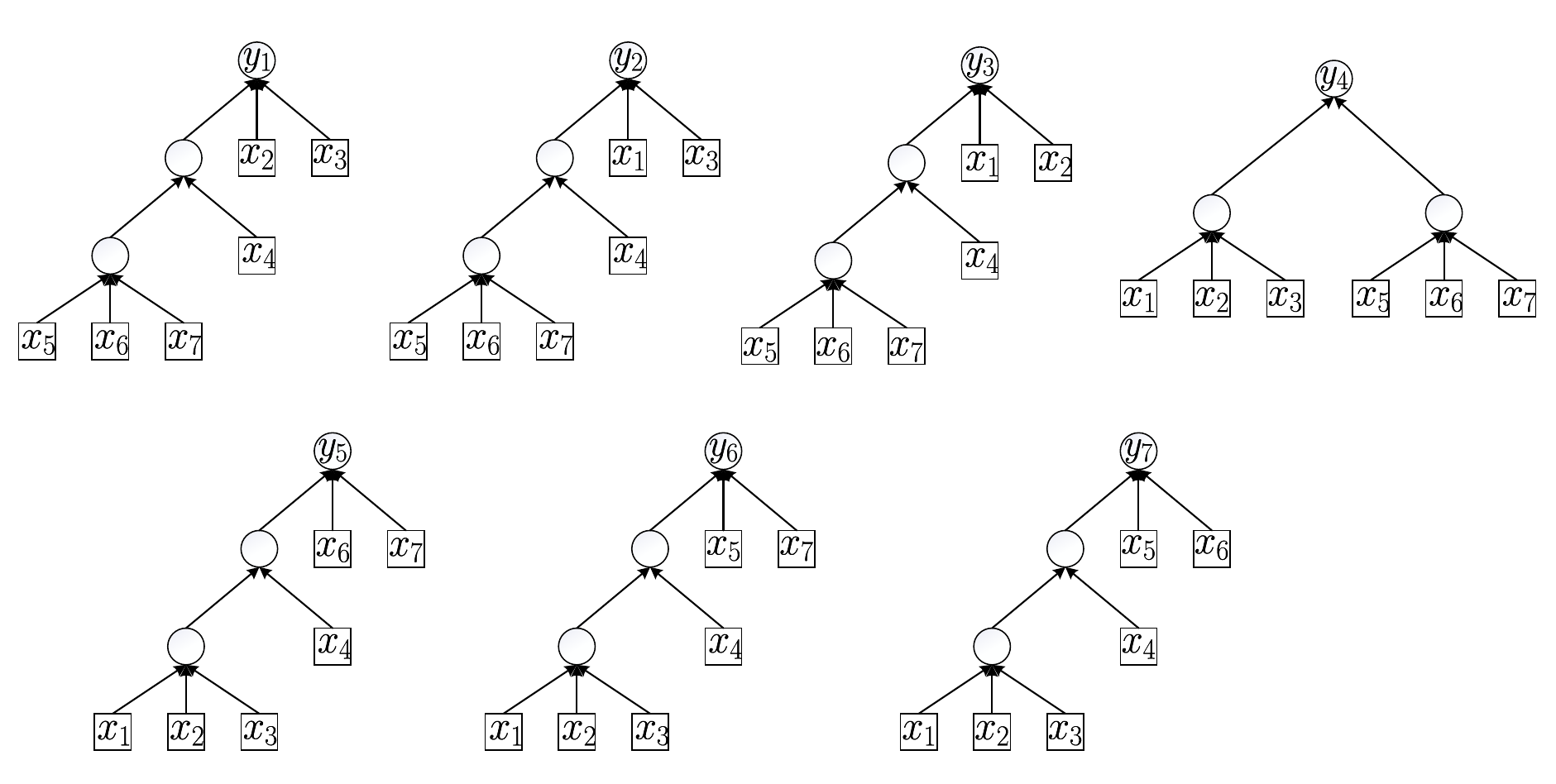}
\caption{The DRTs transformed from the star tree in Fig. \ref{fig: n7_star_tree}(a).}
\label{fig: n7_co_DRTs}
\end{figure*}

\begin{figure}[t!]
\centering
\includegraphics[scale = 0.5]{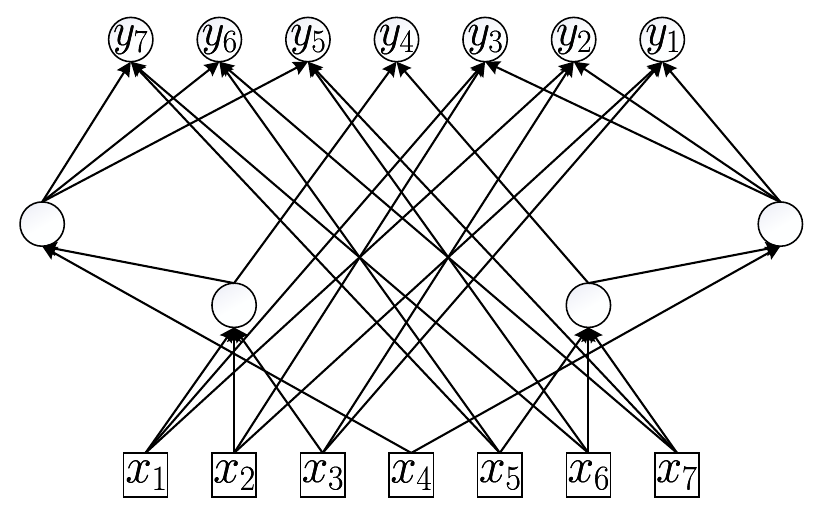}
\caption{A complexity-optimal star-tree-based structure for computation of $\mb{y}$ with $n=7$.}
\label{fig: n7_co_structure}
\end{figure}

Denote $\mc{T}_n$ as the set of all $n$-input star trees.
For example, Fig. \ref{fig: degree3_n7_star_tree} shows the only star tree in $\mc T_3$, and Fig. \ref{fig: n7_star_tree}(a) shows  a star tree in $\mc T_7$.
In  fact, we can obtain a structure based on any star tree.
For instance, consider the star tree in Fig. \ref{fig: n7_star_tree}(a).
First, making $x_1$ as the root and changing undirected edges to directed ones can lead to the DRT in Fig. \ref{fig: n7_star_tree}(b).
Second, removing $x_1$ and labelling the new root as $y_1$ can result in the DRT in Fig. \ref{fig: n7_star_tree}(c) which could be used for computing $y_1$ from $X \setminus \{x_1\}$.
Third, proceed the above steps by replacing $x_1$ and $y_1$ with $x_j$ and $y_j, \forall j \in [2, 7]$, respectively, leading to the seven DRTs in Fig. \ref{fig: n7_co_DRTs}.
Finally, these DRTs can be united (under the union operation $\vee$) to the structure in Fig. \ref{fig: n7_co_structure}.
We formally  describe the construction  from star trees to structures in Construction \ref{construction: star_tree_to_structure}, and use Lemma \ref{lemma: h(T)_str} to state the correctness of the construction.

\begin{construction}\label{construction: star_tree_to_structure}
    Given a star tree $T \in \mc T_n$, return
\[
    h(T) \triangleq \vee_{j \in [n], (a, x_j) \in T} D(a, x_j, T)
\]
 as the constructed star-tree-based structure corresponding to $T$, where $D(a, x_j, T)$ is a DRT defined in Section \ref{section: preliminary}-B.
\end{construction}

All the structures derived from star trees via Construction \ref{construction: star_tree_to_structure} are called star-tree-based structures.
For example, considering $n = 7$ and the star tree in Fig. \ref{fig: n7_star_tree}(a), the DRTs $D(a, x_j, T), \forall j \in [n]$ correspond to the seven DRTs in Fig. \ref{fig: n7_co_DRTs}, and the star-tree-based structure $h(T)$ corresponds to the structure in Fig. \ref{fig: n7_co_structure}.

\begin{lemma} \label{lemma: h(T)_str}
       For any $n \geq 3$ and $T \in \mc T_n$, then $h(T) \in \mc S_n$.
\end{lemma}

\begin{IEEEproof}
    For any $n \geq 3$, $T \in \mc{T}_n$ and $j \in [n]$, $x_j$ is a leaf in $T$.
Let $(a, x_j) \in T$ be the only edge of $x_j$.
$D(a, x_j, T)$ is a DRT with root $y_j$ and leaves $X \setminus \{x_j\}$.
As a result, we have $h(T) \in \mc{S}_n$.
\end{IEEEproof}

According to Lemma \ref{lemma: h(T)_str}, Construction \ref{construction: star_tree_to_structure} is a one-to-one mapping from star trees to star-tree-based structures.
As shown in Fig. \ref{fig: n7_co_structure},  each computation node is either reachable to $y_j$ or reachable from $x_j$, for any $j \in [7]$.
Noting that in any structure, a node reachable from $x_j$ can never be contained in the DRT rooted at $y_j$, the computation nodes in Fig. \ref{fig: n7_co_structure} have been contained in as many DRTs in Fig. \ref{fig: n7_co_DRTs} as possible, indicating that they are fully reused.
In particular, it was proved in \cite{he22aclass}  that for $m = 2$, the star-tree-based structures have the lowest complexity among all the binary-input structures.
This implies that for $m > 2$, star-tree-based structures are likely to  have the near-lowest (and sometimes the lowest) complexity among all structures.
We thus focus on  star-tree-based structures in this section.

\subsection{The Lowest Complexity of Star-Tree-Based Structures}\label{subsection: IV-B}

In this subsection, first, Lemma \ref{lemma: complexity of the star tree} illustrates that the complexity of a star-tree-based structure is totally determined by the degree vector of the corresponding star tree.
Next, Lemma \ref{lemma: degree vector star tree} establishes a necessary and sufficient condition between degree vectors and the existence of a star trees.
Based on these preparations, we finally formulate Problem 1 and then solve it in Theorem \ref{theorem: co-algorithm}, aiming to obtain complexity-optimal star-tree-based structures via  optimizing degree vectors.

\begin{lemma} \label{lemma: complexity of the star tree}
    For any $T \in \mc T_n$,
    %assuming $T$ has $k_i$ degree-$i$ internal nodes for each $i \in [3,m+1]$,
    the complexity of $h(T)$ is
    \begin{equation} \label{eqn: star_tree_complexity}
        c(h(T)) = \sum _{i =3}^{m+1} id_i(T)c_{i-1},
    \end{equation}
    where $d_i(T)$ denotes the number of degree-$i$ nodes in $T$ and $c_{i-1}$  is the complexity factor of $(i-1)$-input nodes.

   % $c(S)=\sum _{i =3}^{m+1} ic_{i-1}$.
\end{lemma}

\begin{IEEEproof}
    It holds that $E(h(T)) =$ $ \cup_{j \in [n], (a, x_j) \in T} E(D(a, x_j, T)) = D(T)$, where the notations can be found in Section \ref{section: preliminary}.
    As a result, we have
    \begin{eqnarray}
        c(h(T))&=&\sum_{a: a \in h(T)}c_{d(a,h(T))}= \sum_{T' \in E(h(T))} c_{d(r(T'),T')}=\sum_{T' \in D(T)} c_{d(r(T'),T')} \notag \\
        &=&\sum _{(a,b)\in T} c_{d(a,D(a,b,T))}=\sum _{(a,b)\in T} c_{d(a,T)-1}=\sum _{i =3}^{m+1} id_i(T)c_{i-1}. \notag
    \end{eqnarray}
\end{IEEEproof}

Recall that the degree vector of $T$ is  $\mb{q} = (d_3(T), d_4(T), \ldots, d_{m+1}(T))$.
Therefore, Lemma \ref{lemma: complexity of the star tree} indicates that $c(h(T))$ is fully determined by $\mb{q}$.
We are interested in optimizing degree vectors to generate complexity-optimal star-tree-based structures.
Since  Construction \ref{construction: star_tree_to_structure} establishes a one-to-one mapping between star trees and star-tree-based structures, the remaining question is that what degree vectors can result in valid $n$-input star trees.
The following lemma answers this question.

\begin{lemma}\label{lemma: degree vector star tree}
For any $n\geq 3$, there exists a star tree $T \in \mc T_n$ with degree vector $\mb q = (q_1, q_2, \ldots, q_{m-1})$ iff
\begin{eqnarray}\label{eqn: star_tree_condition}
    %k_3+\cdots+(m-1)k_{m+1}=n-2.
    2+\sum _{i =1}^{m-1} iq_i=n.
\end{eqnarray}
\end{lemma}

\begin{IEEEproof}
\emph{Necessity:} Let $T \in \mc T_n$.
We prove that the degree vector $\mb{q}$ of $T$ satisfies \eqref{eqn: star_tree_condition}.
    Firstly, for $n=3$, there exists only one star tree as shown in Fig. \ref{fig: degree3_n7_star_tree}, and \eqref{eqn: star_tree_condition} holds obviously.
    Secondly, for $n' \geq 4$, assume \eqref{eqn: star_tree_condition} holds for any $n \in [3, n'-1]$.
    We attempt to prove \eqref{eqn: star_tree_condition} holds for $n=n'$ in the following.
    If $T$ contains only one internal node, \eqref{eqn: star_tree_condition} holds obviously.
    If $T$ contains at least two internal nodes,
     there always exists an internal node $a$ that connects with one internal node and $d(a,T)-1$ leaves.
    Without loss of generality, denote the $d(a,T)-1$ leaves connecting with $a$ as $x_{n'},x_{n'-1},\cdots,x_{n'-d(a,T)+2}$, respectively.
    Then, let $T'=T-x_{n'}-x_{n'-1}-\cdots-x_{n'-d(a,T)+2}$, and relabel $a$ as $x_{n'-d(a,T)+2}$.
    %Return the modified tree as $T'$.
    We can verify that $T' \in \mc T_{n'-d(a,T)+2}$.
    Since \eqref{eqn: star_tree_condition} holds for $n=n'-d(a,T)+2$, we have $2+(\sum _{i =1}^{m-1}iq_i)-(d(a,T)-2)=n'-(d(a,T)-2)$,
    leading to $2+\sum _{i =1}^{m-1} (i-2)q_i=n'$, i.e., \eqref{eqn: star_tree_condition} holds for $n=n'$.
    Therefore, the necessity is proved.
% Since $k_3+\cdots+(m-1)k_{m+1}=n-(d(a)-1)-2$ for $T \in \mc T_{n-(d(a)-1)}$, we have $k_3+\cdots+(m-1)k_{m+1}=n-(d(a)-1)-2$, i.e., \eqref{eqn: star_tree_condition} holds for $T \in \mc T_{n}$.

\emph{Sufficiency:}  Suppose the degree vector $\mb q$ satisfies \eqref{eqn: star_tree_condition}.
We provide Construction \ref{construction: degree vector to star tree} to prove that there always exists a star tree $T \in \mc T_n$ with degree vector $\mb q$.

\begin{construction} \label{construction: degree vector to star tree}
    For a given degree vector $\mb q$, we construct a star tree by the following steps.

Step 1: Select a $q_i>0$.
Then, we can construct an undirected tree $T$, which contains one degree-$(q_i+2)$ internal node and $(q_i+2)$ leaves.
Decrease $q_i$ by 1.

Step 2: If $\mb q= \mb 0$, go to Step 3.
Otherwise, select a $q_i>0$ and select an arbitrary leaf $b \in T$.
Connect $q_i+1$ new leaves to $b$.
We still use $T$ to denote the new tree  and decrease $q_i$ by 1.
Repeat Step 2.

Step 3: Label the leaves in $T$ as $x_1, \ldots, x_{n}$ and keep the internal nodes unlabelled.
Return $T$ as the constructed star tree.
\end{construction}
\end{IEEEproof}

For any $n \geq 3$, there exists at least one degree vector $\mb{q}=(q_1, q_2, \ldots, q_{m-1})$ satisfying \eqref{eqn: star_tree_condition} and to generate star tree via Construction \ref{construction: degree vector to star tree}.
 Note that it may result in multiple star trees, since there are multiple feasible choices of $q_i$ $(1\leq i< m)$ from $\mb q$ in Steps 1 and 2.
Combining with Lemma \ref{lemma: h(T)_str}, $\mb{q}$ corresponds to multiple star-tree-based structures.
Denote $\mc S^\mathrm{star} (\mb q)$ as the set of all the star-tree-based structures derived from  $\mb q$.
Since the complexity of any structure in $\mc S^\mathrm{star} (\mb q)$ is the same by Lemma \ref{lemma: complexity of the star tree},
%and further by Lemma \ref{lemma: degree vector star tree},
it suffices to optimize degree vectors to generate complexity-optimal star-tree-based structures.
Specifically, define $C_{\min}^{\mathrm{star}}(n) \triangleq \min_{T \in \mc{T}_n} c(h(T))$ as the minimum complexity of star-tree-based structures with input size $n$.
Computing $C_{\min}^{\mathrm{star}}(n)$ is equivalent to solving the following optimization problem.
 \begin{eqnarray}
 \textbf{Problem 1:}   &&C_{\min}^{\mathrm{star}}(n)=\min\limits_{\mb q} \, \sum _{i =1}^{m-1} (i+2)q_ic_{i+1} \notag \\
 \,s.t. \quad && 2+\sum _{i =1}^{m-1} iq_i=n.\notag
\end{eqnarray}

The left problem is to solve Problem 1, which  can be done in Algorithm \ref{algorithm: dp_complexity}.
The following theorem illustrates the correctness and time complexity of Algorithm \ref{algorithm: dp_complexity}.

\begin{algorithm}[t!]	
	\caption{Finding optimal solution to Problem 1}
 \label{algorithm: dp_complexity}
	\begin{algorithmic}[1]
		\small
		\REQUIRE   The input size $n$.
		\ENSURE     $C_{\min}^{\mathrm{star}}(n)$.
		%\State
\STATE{Initialize $C_{\min}^{\mathrm{star}}(2)=0$.}
\FOR {$i =3$ \TO $n$}
\STATE{$C_{\min}^{\mathrm{star}}(i)= \min_{t \in [m-1], t \leq i-2} \{C_{\min}^{\mathrm{star}}(i-t)+(t+2)c_{t+1}\}$.}
\ENDFOR
\RETURN {$C_{\min}^{\mathrm{star}}(n)$.}
	\end{algorithmic}
\end{algorithm}

\begin{theorem} \label{theorem: co-algorithm}
    $C_{\min}^{\mathrm{star}}(n)$ can be computed by Algorithm \ref{algorithm: dp_complexity} with time complexity $O(mn)$.
\end{theorem}

\begin{IEEEproof}
In Algorithm \ref{algorithm: dp_complexity},  as shown in Line 1, we have $C_{\min}^{\mathrm{star}}(2)=0$, since $y_1=x_2$ and $y_2=x_1$ for $n=2$, and no computation nodes are required.
Then, for any $i \in [3,n]$, 
assume $\mb q$ is an optimal solution to Problem 1 for $C_{\min}^{\mathrm{star}}(i)$.
For any $t \in [m-1]$ with $q_t>0$, let $\mb q'=\mb q-\mb e_t$, where $\mb e_t$ is a vector with length $m-1$ whose $t$-th entry is 1 and other entries are 0.
Then $\mb q'$ is an optimal solution for $C_{\min}^{\mathrm{star}}(i-t)$, leading to $C_{\min}^{\mathrm{star}}(i)=C_{\min}^{\mathrm{star}}(i-t)+(t+2)c_{t+1}$.
Therefore, $C_{\min}^{\mathrm{star}}(i)=\min_{t \in [m-1], t \leq i-2} \{C_{\min}^{\mathrm{star}}(i-t)+(t+2)c_{t+1}\}$ holds, corresponding to Line 3 of Algorithm \ref{algorithm: dp_complexity}.

Noting that Line 3 in Algorithm \ref{algorithm: dp_complexity} has time complexity $O(m)$ and is carried out $O(n)$ times, the time complexity of Algorithm \ref{algorithm: dp_complexity} is $O(mn)$.
\end{IEEEproof}

By Algorithm \ref{algorithm: dp_complexity}, we are able to determine the lowest complexity $C_{\min}^{\mathrm{star}}(n)$ of star-tree-based structures with input size $n$.
Further by backtracking, we can  obtain an (or all if needed) optimal degree vector $\mb{q}$ which leads to complexity-optimal star-tree-based structures $\mc S^\mathrm{star} (\mb q)$.
However, the structures in $\mc S^\mathrm{star} (\mb q)$ may have different latency.
We are interested in further obtaining the latency-optimal structures in $\mc S^\mathrm{star} (\mb q)$ in the following subsection.

\subsection{The Lowest Latency of Star-Tree-Based Structures for Given Degree Vectors}\label{subsection: IV-C}

In this subsection, first, Lemma \ref{lemma: l(h(T)) = d(T)-1} defines the latency of star trees and shows that it can fully characterize the latency of the corresponding star-tree-based structures.
Then, Lemma \ref{lemma: diameter} illustrates how to compute the latency of star trees.
Based on these preparations, we finally formulate Problem 2 and then solve it in Theorem \ref{theorem: main program}, aiming to obtain  the latency-optimal structures in $\mc S^\mathrm{star} (\mb q)$ via minimizing the latency of star trees corresponding to a given degree vector $\mb{q}$.

\begin{lemma}\label{lemma: l(h(T)) = d(T)-1}
For any star tree $T$, define $$\phi(T) \triangleq \max \left\{\sum_{i=0}^{k} l_{d(v_i,T)-1} :(v_0,v_1,\ldots,v_{k}) \text{ is a simple path in } T\right\}$$ as the latency of $T$.
Then, $l(h(T)) = \phi(T)$.
\end{lemma}

\begin{IEEEproof}
Assume the star tree $T \in \mc T_n$.
Obviously, there exists a simple path $(v_0,v_1,\ldots,v_k) \subseteq T$  iff there exists a path $(v'_0,v'_1,\ldots,v'_{k-1}) \subseteq h(T)$, where $d(v'_i,h(T))=d(v_i,T)-1$ for each $i \in [0,k-1]$.
As a result, we have
$l(h(T))=\max_{(v'_0,v'_1,\ldots,v'_{k-1}) \subseteq h(T)}\{\sum_{i=0}^{k-1}l_{d(v_i,h(T))}\}=\max_{(v_0,v_1,\ldots,v_k) \subseteq T}\{\sum_{i=0}^{k} l_{d(v_i,T)-1}\}=\phi(T)$.
\end{IEEEproof}

As shown in Lemma \ref{lemma: l(h(T)) = d(T)-1}, minimizing the latency of structures derived from the star trees with a given degree vector is equivalent to minimizing the latency of the star trees.  The following lemma shows how to compute the latency of a star tree.

\begin{lemma} \label{lemma: diameter}
For any star tree $T$ with $n \geq 3$, there always exists an edge $(a,b) \in T$ such that
\begin{align}\label{eqn: l(D(a, b, T))}
\left\{
\begin{array}{l}
l(D(a,b,T))-l_{d(a,T)-1}  \leq l(D(b,a,T)),\\
l(D(a,b,T)) > l(D(b,a,T)).
\end{array}
\right.
\end{align}
If $(a,b)$ satisfies \eqref{eqn: l(D(a, b, T))}, we have
\begin{equation}\label{eqn: phiT = l(D(a, b, T)) + }
\phi(T)=l(D(a,b,T))+l(D(b,a,T)).
\end{equation}

\end{lemma}

\begin{IEEEproof}
Suppose $(v_0,v_1,\ldots,v_{k})$ is a simple path in $T$ such that $\sum_{i=0}^{k} l_{d(v_i,T)-1} = \phi(T)$.
There always exists an $i \in [k]$ such that \eqref{eqn: l(D(a, b, T))} holds for $a = v_{i-1}$ and $b = v_{i}$.
This completes the proof of \eqref{eqn: l(D(a, b, T))}.

We are now to prove \eqref{eqn: phiT = l(D(a, b, T)) + }.
Suppose $(a,b) \in T$ is an arbitrary edge satisfying \eqref{eqn: l(D(a, b, T))}.
According to Definition \ref{definition: latency}, $l(D(a,b,T))$ (resp. $l(D(b,a,T))$) denotes the latency of the path in $D(a,b,T)$ (resp. $D(b,a,T)$) from a leaf to $a$ (resp. $b$), and we let this leaf be $x_i$ (resp. $x_j$), without loss of generality.
Then, we have $l(D(a,b,T))+l(D(b,a,T))=\sum_{i=0}^{k} l_{d(v_i,T)-1}$, where $(v_0,v_1,\ldots,v_{k})$ is a simple path in $T$ with $v_0=x_i$ and $v_k=x_j$.
Therefore, we have $\phi(T)\geq l(D(a,b,T))+l(D(b,a,T))$.

To prove  \eqref{eqn: phiT = l(D(a, b, T)) + }, it now suffices to prove $l(T) \leq l(D(a, b, T)) +l(D(b, a, T))$.
Let $x_i$, $x_j$ be two arbitrary leaves in $T$ such that there exists a simple path connecting them and the path's latency is $l(T)$.
If $(x_i \in D(a, b, T)$ and  $x_j \in D(b, a, T))$ or $(x_j \in D(a, b, T)$ and  $x_i \in D(b, a, T))$, we have $l(T) \leq l(D(a, b, T)) +l(D(b, a, T))$.
If $x_i, x_j \in D(a, b, T)$, we have $l(T) \leq 2l(D(a, b, T))  - l_{d(a, T)-1} \leq l(D(a, b, T)) +l(D(b, a, T))$.
If $x_i, x_j \in D(b, a, T)$, we have $l(T) \leq 2l(D(b, a, T))  - l_{d(b, T)-1} < l(D(a, b, T)) +l(D(b, a, T))$.
Therefore,  $l(T) \leq l(D(a, b, T)) +l(D(b, a, T))$.
\end{IEEEproof}

Lemma \ref{lemma: diameter} indicates that, the minimum latency of star trees with degree vector $\mb{q}$ is equal to the  minimum value of $l(D_1) +l(D_2)$, where  $D_1$ and $D_2$ are two DRTs (respectively corresponding to $D(a,b,T)$ and $D(b,a,T)$  in Lemma \ref{lemma: diameter}) satisfy the following three conditions: (i) $l(D_1) - l_{d(r(D_1), D_1)} \leq l(D_2)$, (ii) $l(D_1) > l(D_2)$, and (iii) $d_{i+1}(D_1)+d_{i+1}(D_2)=q_i, \forall i\in [m-1]$, i.e., the summation of $D_1$ and $D_2$'s degree vectors equals $\mb q$.
As a result, computing $L_{\min}^{\mathrm{star}}(\mb{q}) \triangleq \min_{S\in \mc S^{\mathrm{star}}(\mb q)} l(S)$,  the minimum latency of the star-tree-based structures derived from degree vector $\mb{q}$, is equivalent to solving the following optimizing problem.

 \begin{eqnarray}
 \textbf{Problem 2:}   && L_{\min}^{\mathrm{star}}(\mb{q})= \min\limits_{D_1,D_2} \, l(D_1) +l(D_2) \notag \\
 \,s.t. \quad && l(D_1) - l_{d(r(D_1), D_1)} \leq l(D_2), \label{eqn: P2_(1)} \\
 \quad && l(D_1) > l(D_2),\label{eqn: P2_(2)} \\
 \quad && d_{i+1}(D_1)+d_{i+1}(D_2)=q_i, \forall i\in [m-1].\label{eqn: P2_(3)}
\end{eqnarray}

To this end, for any degree vector $\mb u$ and positive integer $t$, define $\tau(\mb u, t) = \min_{D'_1, ..., D'_t}$ $ \max \{l(D'_1), ..., l(D'_t)\}$,  where $D'_1, ..., D'_t$ are DRTs and the summation of their degree vectors equals $\mb u$.
Here, $\tau(\mb u,t)$ returns the lowest latency of the substructures that have the degree vector $\mb u$ and consist of $t$ disjoint DRTs.
It will be computed later in Algorithm \ref{algorithm: subprogram}.
For any two degree vectors $\mb u$, $\mb u' \in \mathbb{Z}_{+}^{m-1}$, we say $\mb u \preceq \mb u'$ iff for each $i \in [m-1]$, $u_i \leq u_i'$.
Then, we can solve Problem 2 by Algorithm \ref{algorithm: main program}.

\begin{algorithm}[H]	
	\caption{Finding optimal solution to Problem 2}
	\label{algorithm: main program}
	\begin{algorithmic}[1]
		\small
		\REQUIRE    $\mb q$.
		\ENSURE  $L_{\min}^{\mathrm{star}}(\mb{q})$
\STATE \ti{\%$\mb{u}$ and $\mb{q} - \mb{u}$ are the degree vectors of $D_1$ and $D_2$, respectively}
\FOR{each $\mb u \preceq \mb q$}
    \STATE \ti{\%The degree of $D_1$'s root is $d(r(D_1), D_1) = i+1$}
		\FOR {$i =1$ \TO $m-1$}
\IF{$u_{i} \geq 1$}
\STATE \ti{\%As $\mb{u}$ and $i$ are given, the minimum values of $l(D_1)$ and $l(D_2)$ are $\tau(\mb u-\mb e_{i},i+1) +l_{i+1}$ and $\tau(\mb q-\mb u,1)$, respectively}
\IF{$\tau(\mb u-\mb e_{i},i+1) \leq \tau(\mb q-\mb u,1)$ (i.e., Eq. \eqref{eqn: P2_(1)}) and $\tau(\mb u-\mb e_{i},i+1) +l_{i+1}> \tau(\mb  q-\mb u,1) $ (i.e., Eq. \eqref{eqn: P2_(2)})}

\STATE{$l=  \tau(\mb u-\mb e_{i},i+1) +l_{i+1}+\tau(\mb q-\mb u,1)$. \ti{\%The minimum value of $l(D_1) + l(D_2)$ for given $\mb{u}$ and $i$}}

\STATE{$L_{\min}^{\mathrm{star}}(\mb{q})=\min \{ L_{\min}^{\mathrm{star}}(\mb{q}), l\}$.}

\ENDIF
\ENDIF
\ENDFOR
\ENDFOR

\RETURN $L_{\min}^{\mathrm{star}}(\mb{q})$.

	\end{algorithmic}
\end{algorithm}

\begin{theorem}\label{theorem: main program}
    Given a degree vector $\mb q$ and $\tau(\mb u, t)$, $\forall \mb u \leq \mb q, t \in [m]$, Algorithm \ref{algorithm: main program} can compute $L_{\min}^{\mathrm{star}}(\mb{q})$ with time complexity  $O(m  \prod_{i=1}^{m-1} (q_i+1))$.
\end{theorem}

\begin{IEEEproof}
Let $\mb u$ be the degree vector  of $D_1$ and $i+1$ be the degree of $D_1$'s root.
According to constraint \eqref{eqn: P2_(3)}, the degree vector of $D_2$ must be $\mb q-\mb u$.
Therefore, as shown in Lines 2--13, by traversing $\mb u \preceq \mb q$ and $i$ from 1 to $m-1$, we can enumerate all the feasible solutions and find the one with lowest $l(D_1)+l(D_2)$.
Specifically, given $\mb u$ and $i$,  the minimum latency of $D_1$ and $D_2$ are $\tau(\mb q-\mb e_i,i+1)+l_{i+1}$ and $\tau(\mb q-\mb u,1)$ respectively, and the conditions in Line 7 correspond to the constraints \eqref{eqn: P2_(1)} and \eqref{eqn: P2_(2)}.

In Algorithm \ref{algorithm: main program}, Line 8  is carried out $O(m  \prod_{i=1}^{m-1} (q_i+1))$ times, each of which has time complexity $O(1)$.
Therefore, the time complexity of Algorithm \ref{algorithm: main program} is $O(m  \prod_{i=1}^{m-1} (q_i+1))$.
\end{IEEEproof}

At this point, the left problem is to compute $\tau(\mb u, t)$, $\forall \mb u \leq \mb q, t \in [m]$.
We complete this task in Algorithm \ref{algorithm: subprogram}.

\begin{algorithm}[H]	
%\begin{algorithm}[t!]
    %\captionsetup{font=small}
	%\renewcommand{\thealgorithm}{1.1}
	\caption{Computation of $\tau(\mb u,t), \forall \mb u \preceq \mb q, t \in [m].$}
	\label{algorithm: subprogram}
	\begin{algorithmic}[1]
		\small
		\REQUIRE    $\mb q$.
		\ENSURE     $\tau(\mb u,t), \forall \mb u \preceq \mb q, t \in [m].$
  \FOR {$t =1$ \TO $m$}
\STATE{$\tau(\mb 0,t)=0$.}
\ENDFOR

 \FOR{each $\mb u \preceq \mb q$  such that $||\mb u||_1$ is non-decreasing }

\FOR {$t =1$ \TO $m$}
\STATE \ti{\%Suppose $G$ is an arbitrary substructure that has degree vector $\mb u$ and consists of $t$ disjoint DRTs}
\IF{$t==1$}
\STATE \ti{\%
$G$ is a DRT.
Fixing the degree of its root as $d(r(G),G)=i+1$,
the graph $G-r(G)$ has degree vector $\mb u-\mb e_i$ and consists of $i+1$ disjoint DRTs }
\STATE{$\tau(\mb u,t)=\min_{i \in [m-1], u_i>0}\{\tau(\mb u-\mb e_{i},i+1)+l_{i+1}\}$.}
\ELSE
\STATE \ti{\%
$G$ consists of $t$ disjoint DRTs, named $D_1,D_2,\ldots,D_t$.
Fixing the degree vector of $D_1$ as $\mb u'$,
the graph $G\setminus D_1$ has degree vector $\mb u-\mb u'$ and  consists of $t-1$ disjoint DRTs }
\STATE{$\tau(\mb u,t)=\min_{\mb u' \preceq \mb u} \max\{\tau(\mb u',1), \tau(\mb u-\mb u',t-1)\}$.}
\ENDIF
\ENDFOR
\ENDFOR
\RETURN {$\tau(\mb u,t), \forall \mb u \preceq \mb q, t \in [m].$}

	\end{algorithmic}
\end{algorithm}

\vspace{0.75cm}

\begin{lemma}\label{lemma:subprogram}
    Given a degree vector $\mb q$, then $\tau(\mb u, t), \forall \mb u \preceq \mb q, t \in [m]$ can be computed by Algorithm \ref{algorithm: subprogram} with time complexity $O(m  \prod_{i=1}^{m-1}\frac{(q_i+1)(q_i+2)}{2})$.
\end{lemma}

\begin{IEEEproof}
Firstly, as shown in Line 2,  for any $t\in [m]$, we  have $\tau(\mb 0,t)=0$,  since there exist no computation nodes in the graph.
Then, for any degree vector $\mb u \preceq \mb q$ and $t \in [m]$, suppose $G$ is an arbitrary substructure that has degree vector $\mb u$ and consists of $t$ disjoint DRTs.
If $t=1$, the substructure $G$ is a DRT.
Fixing the degree of its root as $d(r(G),G)=i+1$,
the graph $G-r(G)$ has degree vector $\mb u-\mb e_i$ and consists of $i+1$ disjoint DRTs.
Therefore, when fixing $d(r(G),G)=i+1$, the minimum latency of $G$ is $\tau(\mb u-\mb e_i,i+1) + l_{i+1}$.
As shown in Line 7, by traversing all the possible $i$, we can find the minimum latency of $G$:
\begin{equation} \label{eqn: case t=1}
    \tau(\mb u,t)=\min_{i \in [m-1], u_i>0}\{\tau(\mb u-\mb e_{i},i+1)+l_{i+1}\}.
\end{equation}
If $t>1$, the substructure $G$ consists of $t$ disjoint DRTs, named $D_1,D_2,\ldots,D_t$.
Fixing the degree vector of $D_1$ as $\mb u'$,
the graph $G \setminus D_1$ has degree vector $\mb u-\mb u'$ and  consists of $t-1$ disjoint DRTs.
Therefore, when fixing the degree vector of $D_1$ as $\mb u'$, the minimum latency of $G$ is $\max\{\tau(\mb u',1), \tau(\mb u-\mb u',t-1)\}$.
As shown in Line 7, by traversing all the possible $\mb u'$, we can find the minimum latency of $G$:
\begin{equation} \label{eqn: case t>1}
    \tau(\mb u,t)=\min_{\mb u' \preceq \mb u} \max\{\tau(\mb u',1), \tau(\mb u-\mb u',t-1)\}.
\end{equation}
As shown in Lines 4--10, we enumerate all the possible $\mb u$ and $t$ and compute $\tau(\mb u,t)$.
Here, we enumerate $\mb u$ such that $||\mb u||_1$ is non-decreasing and enumerate $t$ from 1 to $m$, so that
when computing $\tau(\mb u,t)$, the required values of $\tau(\mb u',t')$ have been already computed.
Therefore, given $\mb q$, Algorithm \ref{algorithm: subprogram} can obtain $\tau(\mb u, t), \forall \mb u \preceq \mb q, t \in [m]$.

In the following, we investigate the time complexity of Algorithm \ref{algorithm: subprogram}.
In Line 4, we enumerate $\mb u \preceq \mb q$.
In Line 5,  we enumerate $t$ from $1$ to $m$.
And in Line 9, we enumerate $\mb u' \preceq \mb u$.
Thus, the number of computations is $\sum_{\mb u \preceq \mb q} m|\{\mb u': \mb u' \preceq \mb u\}|$.
We can simplify it as follows:
\begin{eqnarray*}
    &\,&\sum_{\mb u \preceq \mb q} m|\{\mb u': \mb u' \preceq \mb u\}| \\
    &=&m\sum_{\mb u \preceq \mb q} |\{\mb u': \mb u' \preceq \mb u\}| \\
    &=& m\sum_{ u_1 = 0}^{q_1}\sum_{  (u_2,\ldots,u_{m-1}) \preceq  (q_2,\ldots,q_{m-1})} |\{\mb u': \mb u' \preceq \mb u\}| \\
    &=& m\sum_{ u_1 = 0}^{q_1}\sum_{  (u_2,\ldots,u_{m-1}) \preceq  (q_2,\ldots,q_{m-1})} |\{ (u'_1,\ldots, u'_{m-1}): u'_1 \in [0,u_1], (u'_2,\ldots,u'_{m-1}) \preceq  (u_2,\ldots,u_{m-1})\}| \\
            \end{eqnarray*}
    \begin{eqnarray*}
    &=& m\sum_{ u_1 = 0}^{q_1}((u_1+1)\sum_{  (u_2,\ldots,u_{m-1}) \preceq  (q_2,\ldots,q_{m-1})} |\{ (u'_2,\ldots, u'_{m-1}):  (u'_2,\ldots,u'_{m-1}) \preceq  (u_2,\ldots,u_{m-1})\}|) \\
    &=& m\frac{(q_1+1)(q_1+2)}{2}\sum_{  (u_2,\ldots,u_{m-1}) \preceq  (q_2,\ldots,q_{m-1})} |\{ (u'_2,\ldots, u'_{m-1}):  (u'_2,\ldots,u'_{m-1}) \preceq  (u_2,\ldots,u_{m-1})\}| \\
    &=& m\frac{(q_1+1)(q_1+2)}{2}\frac{(q_2+1)(q_2+2)}{2}\\
    &\,&\sum_{  (u_3,\ldots,u_{m-1}) \preceq  (q_3,\ldots,q_{m-1})} |\{ (u'_3,\ldots, u'_{m-1}):  (u'_3,\ldots,u'_{m-1}) \preceq  (u_3,\ldots,u_{m-1})\}| \\
    &=& m\prod_{i=1}^{m-1}\frac{(q_i+1)(q_i+2)}{2}.
\end{eqnarray*}
The proof is completed.
\end{IEEEproof}

By Algorithms \ref{algorithm: main program} and \ref{algorithm: subprogram}, we are able to determine the lowest latency $L_{\min}^{\mathrm{star}}(\mb{q})$ of the star-tree-based structures $\mc S^{\mathrm{star}}(\mb q)$ derived from degree vector $\mb{q}$.
Further by backtracking, we can
find the corresponding latency-optimal star tree $T$ with degree vector $\mb{q}$, and $T$ can be further used to generate a latency-optimal star-tree-based structure $S$ from $\mc S^{\mathrm{star}}(\mb q)$ via Construction \ref{construction: star_tree_to_structure}.
Note that $\mb{q}$ can be any valid degree vector.
However, if $\mb q$ is the uniquely optimal solution to Problem 1,
$S$ has the lowest latency among all the complexity-optimal star-tree-based structures.
On the other hand, if Problem 1 has multiple optimal solutions, we can enumerate them to find the latency-optimal ones among complexity-optimal star-tree-based structures.

\section{Isomorphic-DRT-Based Structures}\label{section: Isomorphic-DRT-Based Structures}

In this section, we construct structures for the scenario where latency has a higher priority than complexity.
To this end, we focus on a class of structures, called isomorphic-DRT-based structures.
Since there always exist isomorphic-DRT-based structures (or structures derived from them) achieving the lowest latency, which is explicated in Lemma \ref{lemma: DRT latency}.
Section \ref{subsection: V-A} proposes a class of  DRTs, called isomorphic DRTs, and further provides a construction from isomorphic DRTs to isomorphic-DRT-based structures.
Next, Section \ref{subsection: V-B} presents a construction from type vectors to isomorphic-DRTs.
Here, the type vector is used to characterize an isomorphic DRT, each component of which is the number of levels that contain nodes with the corresponding degree.
We further illustrate that given a type vector, the latency of its corresponding isomorphic-DRT-based structures is uniquely determined.
Therefore, Theorem \ref{theorem: lo-algoritm} is proposed to find the type vectors that lead to the latency-optimal isomorphic-DRT-based structures.
Finally, Section \ref{subsection: V-C}  derives Construction \ref{construction:  latency optimal} and Theorem \ref{theorem: complexity of isomorphic DRT} to find a complexity-optimal  isomorphic-DRT-based structure $S$ corresponding to a given type vector $\mb{w}$.
As a result, if $\mb{w}$ is the uniquely optimal type vector obtained from Theorem \ref{theorem: lo-algoritm},  then $S$ is a latency-optimal isomorphic-DRT-based structure which also has the lowest complexity among all latency-optimal isomorphic-DRT-based structures.

\subsection{Isomorphic DRTs and Isomorphic-DRT-Based Structures}\label{subsection: V-A}

In this subsection, we propose a class of DRTs, called isomorphic DRTs, in Definition \ref{def: isomorphic DRT}.
 Then, Construction \ref{construction: isomorphic DRT to structure} and Lemma \ref{lemma: DRT_to_structure} show that, based on any isomorphic DRT $D$, we can obtain a structure, which is called an  isomorphic-DRT-based structure.
 Finally, Lemma \ref{lemma: DRT latency} illustrates that it suffices to investigate the isomorphic-DRT-based structures when the latency of structures is of interest.

\begin{definition} [isomorphic DRT]\label{def: isomorphic DRT}
    We refer to a DRT $D$ as an  isomorphic DRT iff
    %For any two nodes $a,b     \in T$, we have $E(a,T)$ and $E(b,T)$ are isomorphic
for any internal node $a$ in $D$, all the subtrees rooted at $a$'s children are isomorphic.
\end{definition}

Denote $\mc{D}_n$ as the set of all isomorphic DRTs with exactly $n$-leaves.
Fig. \ref{fig: n7_non-optimal_DRTs} shows an example of isomorphic DRTs in $\mc{D}_6$.
In an  isomorphic DRT $D$, the nodes in a same level have the same degree (number of incoming edges).
For the sake of convenience, for any $i \in [2, m]$, we refer to a level of $D$ by type-$i$ level if it contains degree-$i$ nodes.
Moreover, we define the type vector of $D$ by $\mb{w} = (w_1, w_2, \ldots, w_{m-1}) \in \mbb{Z}_+^{m-1}$, where $w_i, i \in [m-1]$ is the number of type-$(i+1)$ levels in $D$.

In the following, we propose a construction  from isomorphic DRTs to structures in Construction \ref{construction: isomorphic DRT to structure}, and use Lemma \ref{lemma: DRT_to_structure} to state the correctness of the construction.

\begin{construction}\label{construction: isomorphic DRT to structure}
 For any $n \geq 3$ and an  isomorphic DRT $D \in \mc D_{n-1}$, we construct the structure by the following steps:

Step 1: For any $j \in [n]$, let $D_j$ be a copy of $D$, and label all the leaves of $D_j$ as $X \setminus \{x_j\}$ respectively.

Step 2: Return $S=\vee_{j\in[n]}D_j$ as the constructed structure.
\end{construction}

\begin{lemma} \label{lemma: DRT_to_structure}
       For any $n \geq 3$ and $D \in \mc D_{n-1}$, supposing that $S$ is
 a structure derived from $D$ via Construction \ref{construction: isomorphic DRT to structure}, then $S \in \mc S_{n}$.
\end{lemma}

\begin{IEEEproof}
    It holds obviously, according to Definitions \ref{definition: structure} and \ref{construction: isomorphic DRT to structure}.
\end{IEEEproof}
%not all the n

All the structures that can be derived from isomorphic DRTs via Construction \ref{construction: isomorphic DRT to structure} are called isomorphic-DRT-based structures.
Fig. \ref{fig: S7_structure_Ueng2017} shows an example of isomorphic-DRT-based structure corresponding to the isomorphic DRTs given in Fig. \ref{fig: n7_non-optimal_DRTs}.
Note that the resultant structure of Construction \ref{construction: isomorphic DRT to structure} is generally not  unique for a given isomorphic DRT $D$, since we may label the leaves of $D_j$ in different ways.
For convenience, denote $\mc S^{\mathrm{isom}}(D)$ as the set of all possible isomorphic-DRT-based structures derived from $D$ via Construction \ref{construction: isomorphic DRT to structure}.
 
In the following, we illustrate that the minimum latency of the structures for given $n$ can be connected to the minimum latency of isomorphic-DRT-based structures with input size $n'\geq n$.

\begin{lemma}\label{lemma: DRT latency}
    For any $S \in \mc S_n$, there always exists an  isomorphic-DRT-based structure $S' \in \mc S_{n'}$ satisfying that $n'\geq n$ and $l(S') \leq l(S)$.
\end{lemma}

\begin{IEEEproof}
For a given structure $S\in \mc S_n$, we attempt to construct an  isomorphic-DRT-based structure $S'$ satisfying that $n'\geq n$ and $l(S') \leq l(S)$ by the following steps.

Step 1: Let $D$ denote the DRT $E(y_1,S)$ and then unlabel $D$.
Set $i=1$.

Step 2: %Let $A_i$ be the set of the nodes in the $i$-th level.
 Let $A_i$ be the set of nodes in the $i$-th level of $D$.
If all the nodes in $A_i$ are leaves, go to Step 3.
Otherwise, select a node $a \in  A_i$, such that the number of leaves of $E(a,D)$ is equal to or greater than $E(b,D), \forall b \in A_i$.
Then, for each $b \in A_i \setminus \{a\}$, modify the $E(b,D)$, so that $E(b,D)$  is isomorphic to $E(a,D)$.
Increase $i$  by 1 and then repeat Step 2.

Step 3: 
Return $S'$ derived from $D$ via Construction \ref{construction: isomorphic DRT to structure}.

In Step 2, the number of $D$'s leaves is either increased or unchanged, and the latency of $D$ is at most $l(S)$.
As a result, $S'$ is an  isomorphic-DRT-based structure satisfying that $n'\geq n$ and $l(S') \leq l(S)$.
The proof is completed.
\end{IEEEproof}

\begin{figure}[t]
\centering
\includegraphics[scale = 0.5]{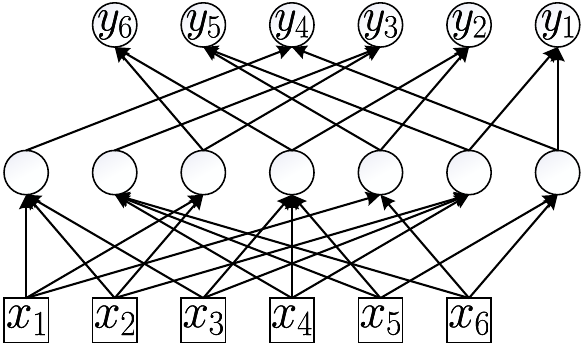}
\caption{A structure in $\mc S_6$ which is generated from that in Fig. \ref{fig: n7_structure} by removing  nodes $x_7$ and $y_7$.}
\label{fig: n6_structure_modified}
\end{figure}

According to Lemma \ref{lemma: DRT latency}, it suffices to investigate the latency-optimal isomorphic-DRT-based structures whose number of leaves is at least $n$ when the minimum latency of $S \in \mc S_n$ is of interest.
In particular, if we can first construct a latency-optimal isomorphic-DRT-based structure $S' \in \mc{S}_{n'}$ with $n' \geq n$,  then properly we remove certain nodes (including $x_{n+1},..., x_{n'}$, $y_{n+1},..., y_{n'}$) from $S'$ to obtain a latency-optimal structure $S \in \mc S_n$, where
$l(S) = l(S')$ by Lemma \ref{lemma: DRT latency} and by the fact that removing nodes from $S'$ does not increase latency.
For instance, we can  initially construct the latency-optimal isomorphic-DRT-based structure $S' \in \mc S_7$ shown in Fig. \ref{fig: n7_structure}, and then remove the nodes $x_7$ and $y_7$ from $S'$ to obtain the latency-optimal $S \in \mc{S}_6$ shown in  Fig. \ref{fig: n6_structure_modified}.

According to the above analysis,  we only focus on the latency and complexity of isomorphic-DRT-based structures in the rest of this section,.

\subsection{The Lowest Latency of Isomorphic-DRT-Based Structures}\label{subsection: V-B}

In this subsection, first, Lemma \ref{lemma: latency of the isomorphic DRT} investigates the latency of isomorphic-DRT-based structures and find that given a type vector, the latency of its corresponding isomorphic-DRT-based structures is uniquely determined.
Next, Construction \ref{construction: type vector to isomorphic DRT} provides a method to obtain isomorphic DRTs for given type vectors.
Based on these preparations, we finally formulate Problem 3 and solve it in Theorem \ref{theorem: lo-algoritm}, aiming to obtain latency-optimal  isomorphic-DRT-based structures via  optimizing type vectors.

\begin{lemma} \label{lemma: latency of the isomorphic DRT}
    For any $n\geq 3$, $D \in \mc D_{n-1}$ and $S \in \mc S^{\mathrm{isom}}(D)$,
    the latency of $S$ is
    \begin{equation} \label{eqn: star_tree_complexity}
        l(S) = l(D) = \sum_{i=1}^{m-1}w_i l_{i+1},
    \end{equation}
    where $w_i$ denotes the number of type-$(i+1)$ levels in $D$ and $l_{i+1}$  is the latency factor of $(i+1)$-input nodes.

   % $c(S)=\sum _{i =3}^{m+1} ic_{i-1}$.
\end{lemma}

\begin{IEEEproof}
According to Definition \ref{def: isomorphic DRT}, for any $i,j \in [n+1]$ and $i \neq j$, the path from $x_i$ to $y_j$ has the latency $\sum_{i=1}^{m-1}w_i l_{i+1}$.
Noting Definition \ref{definition: complexity latency}, the proof is completed.
\end{IEEEproof}

%Recall that the degree vector of $T$ is  $\mb{q} = (d_3(T), d_4(T), \ldots, d_{m+1}(T))$.
According to Lemma \ref{lemma: latency of the isomorphic DRT}, the latency $l(S)$ is fully determined by the type vector $\mb{w} = (w_1, w_2, \ldots, w_{m-1})$.
We are interested in optimizing type vectors to generate latency-optimal isomorphic-DRT-based structures.
Since Construction \ref{construction: isomorphic DRT to structure} can generate isomorphic-DRT-based structures from isomorphic DRTs, the remaining question is that what type vectors can result in valid $(n-1)$-input isomorphic DRTs.
The following lemma answers this question.

\begin{lemma}\label{lemma: number of leaves of isomorphic DRT}
For any $n\geq 3$, there exists an  isomorphic DRT $D \in \mc D_{n-1}$ with type vector $\mb w = (w_1, w_2, \ldots, w_{m-1})$ iff
\begin{eqnarray}\label{eqn: isomorphic_DRT_condition}
    %k_3+\cdots+(m-1)k_{m+1}=n-2.
    \prod_{i=1}^{m-1} (i+1)^{w_i}=n-1.
\end{eqnarray}
\end{lemma}

\begin{IEEEproof}
    \emph{Necessity:} According to Definition \ref{def: isomorphic DRT}, the necessity holds obviously.

    \emph{Sufficiency:}  Suppose the type vector $\mb w$ satisfies \eqref{eqn: isomorphic_DRT_condition}.
We provide Construction \ref{construction: type vector to isomorphic DRT} to prove that there always exists an  isomorphic DRT $D \in \mc D_{n-1}$ with type vector $\mb w$.

\begin{construction} \label{construction: type vector to isomorphic DRT}
    For a given type vector $\mb w$, we construct an  isomorphic DRT by the following steps.

Step 1: Select a $w_i>0$.
Then, we can construct a DRT $D$, which contains one degree-$(i+1)$ internal node and $(i+1)$ leaves.
Decrease $w_i$ by 1.

Step 2: If $\mb w= \mb 0$, return $D$ as the constructed isomorphic DRT.
Otherwise, select a $w_i>0$ and replace each leave in $D$ with a subtree that contains one degree-$(i+1)$ internal node and $(i+1)$ leaves.
Repeat Step 2.
%Step 3: Return $D$ as the constructed isomorphic DRT.
\end{construction}
\end{IEEEproof}

Note that for some $n \geq  3$, there may exist no type vector satisfying \eqref{eqn: isomorphic_DRT_condition} such that $\mc{D}_{n - 1} = \emptyset$.
However, as discussed at the end of Section \ref{subsection: V-A}, we only need to focus on those $n$ such that \eqref{eqn: isomorphic_DRT_condition} can be satisfied and $\mc{D}_{n - 1} \neq \emptyset$.
Further recall that  a type vector can generate multiple isomorphic DRTs via Construction \ref{construction: type vector to isomorphic DRT} and  an  isomorphic DRT can also generate multiple isomorphic-DRT-based structures via Construction \ref{construction: isomorphic DRT to structure}.
As a result, a type vector corresponds to multiple isomorphic-DRT-based structures.
Denote $\mc S^{\mathrm{isom}} (\mb w)$ as the set of all the isomorphic-DRT-based  structures derived from degree vector $\mb w$.
Since the latency of any structure in $\mc S^{\mathrm{isom}} (\mb w)$ is the same by Lemma \ref{lemma: latency of the isomorphic DRT}, it suffices to optimize type vectors to generate latency-optimal isomorphic-DRT-based structures.
Specifically, define $L_{\min}^{\mathrm{isom}}(n-1) \triangleq \min_{D \in \mc{D}_{n-1}} l(D)$ for $\mc{D}_{n-1} \neq \emptyset$, which is the minimum latency of isomorphic DRTs with input size $n-1$ as well as the minimum latency of isomorphic-DRT-based structures with input size $n$.
Computing $L_{\min}^{\mathrm{isom}}(n-1)$ is equivalent to solving the following optimization problem.

 \begin{eqnarray}
 \textbf{Problem 3:}   &&L_{\min}^{\mathrm{isom}}(n-1) = \min\limits_{\mb w} \, \sum_{i=1}^{m-1}w_i l_{i+1}\notag \\
 \,s.t. \quad && \prod_{i=1}^{m-1} (i+1)^{w_i} = n-1. \label{eqn: Problem 3}
\end{eqnarray}

The left problem is to solve Problem 3, which has been done in Algorithm \ref{algorithm: dp_latency}.
The following theorem illustrates the correctness and time complexity of Algorithm \ref{algorithm: dp_latency}.

\begin{algorithm}[t!]	
	\caption{Finding optimal solutions to Problem 3}
 \label{algorithm: dp_latency}
	\begin{algorithmic}[1]
		\small
		\REQUIRE   The input size $n$.
		\ENSURE     $L_{\min}^{\mathrm{isom}}(n-1)$.
		%\State
\STATE{Initialize $L_{\min}^{\mathrm{isom}}(1)=0$.}
\FOR {$i =2$ \TO $n-1$}
\STATE{$L_{\min}^{\mathrm{isom}}(i) = \min_{t \in [2, m] \text{~is a factor of~} i} \{L_{\min}^{\mathrm{isom}}(i/t ) + l_t\}$.}
\ENDFOR
\RETURN {$L_{\min}^{\mathrm{isom}}(n-1)$.}
	\end{algorithmic}
\end{algorithm}

\begin{theorem}\label{theorem: lo-algoritm}
    $L_{\min}^{\mathrm{isom}}(n-1)$ can be computed by Algorithm \ref{algorithm: dp_latency} with time complexity  $O(mn)$.
\end{theorem}

\begin{IEEEproof}
    In Algorithm \ref{algorithm: dp_latency},  as shown in Line 1, we have $L_{\min}^{\mathrm{isom}}(1)=0$, since no computation nodes are required.
Then, for any $i \in [2,n-1]$, assume $\mb w$ is an optimal solution to Problem 3 for $L_{\min}^{\mathrm{isom}}(i)$.
For any integer $ t \in [2, m]$ satisfying $w_{t-1}>0$, then $t$ must be a factor of $i$.
Let $\mb w'=\mb w-\mb e_{t-1}$, where $\mb e_{t-1}$ is a vector with length $m-1$ whose $(t-1)$-th entry is 1 and other entries are 0.
Then we have $\mb w'$ is an optimal solution for $L_{\min}^{\mathrm{isom}}( i/t )$, leading to $L_{\min}^{\mathrm{isom}}(i)=L_{\min}^{\mathrm{isom}}( i/t )+l_t$.
Therefore, $L_{\min}^{\mathrm{isom}}(i) = \min_{t \in [2, m] \text{~is a factor of~} i} \{L_{\min}^{\mathrm{isom}}(i/t ) + l_t\}$ holds, corresponding to Line 3 of Algorithm \ref{algorithm: dp_latency}.

Noting that Line 3 in Algorithm \ref{algorithm: dp_latency} has time complexity $O(m)$ and is carried out $O(n)$ times, the time complexity of Algorithm \ref{algorithm: dp_latency} is $O(mn)$.
\end{IEEEproof}

By Algorithm \ref{algorithm: dp_latency}, we are able to determine the lowest latency $L_{\min}^{\mathrm{isom}}(n-1)$ of isomorphic-DRT-based structures  with input size $n$.
Further by backtracking, we can  obtain an (or all if needed) optimal type vector $\mb{w}$ which leads to latency-optimal isomorphic-DRT-based structures $\mc S^\mathrm{isom} (\mb w)$.
However, the structures in $\mc S^\mathrm{isom} (\mb w)$ may have different complexity.
We are interested in further obtaining the complexity-optimal structures in $\mc S^\mathrm{isom} (\mb w)$ in the following subsection.

\subsection{The Lowest Complexity of Isomorphic-DRT-Based Structures for Given Type Vectors}\label{subsection: V-C}

In this subsection, for a given degree vector $\mb{w}$, we first generate an  isomorphic-DRT-based  structure $S \in \mc S^\mathrm{isom} (\mb w)$ via  Construction \ref{construction:  latency optimal}, and then prove that $S$ has the lowest complexity among  all isomorphic-DRT-based structures $\mc S^\mathrm{isom} (\mb w)$ in Theorem \ref{theorem: complexity of isomorphic DRT}.

\begin{construction}\label{construction:  latency optimal}

For a given type vector $\mb w$, we construct a structure $S$ by the following steps.

Step 1: Obtain an  isomorphic DRT $D$ from $\mb w$ via Construction \ref{construction: type vector to isomorphic DRT}.

Step 2: Obtain an  isomorphic-DRT-based structure $S$ from $D$ via Construction \ref{construction: isomorphic DRT to structure}, where for each $j \in [n]$, label leaves of $D_j$ from left to right in a consecutive way:  $x_{j+1}, \ldots, x_n, x_1, \ldots, x_{j-1}$, where $x_n$ and $x_1$ are considered as consecutive to each other.
Return $S$ as the constructed structure.
\end{construction}

Fig. \ref{fig: n7_structure} shows a structure $S$ obtained from  Construction \ref{construction: latency optimal} for $n=7$, $m=3$ and $\mb w=(1,1)$.
The following theorem states that $S$ does have the lowest complexity among $\mc S^\mathrm{isom} (\mb w)$.

\begin{theorem}\label{theorem: complexity of isomorphic DRT}
For a given type vector $\mb w$  and $n = 1 +\prod_{i=1}^{m-1} (i+1)^{w_i}$, let $S$ be the structure obtained from Construction \ref{construction: latency optimal}.
Then, $S \in \mc S^\mathrm{isom} (\mb w)$ and $c(S) = \sum_{i=1}^{m-1} n w_i c_{i+1} =  \min_{S' \in \mc S^{\mathrm{isom}} (\mb w)}c(S')$.
\end{theorem}

\begin{IEEEproof}
Let $S'\in \mc S^{\mathrm{isom}}(\mb w)$.
We first prove
\begin{equation}
    c(S') \geq \sum_{i=1}^{m-1}nw_ic_{i+1}.\label{eqn: T4}
\end{equation}
For any $j \in [n]$, $E(y_j, S')$ must be an  isomorphic DRT of height $\zeta$, where $\zeta=\sum_{i=1}^{m-1} w_i$.
For $i \in [0,\zeta]$, let $H_i$ be the $i$-th level in $S'$, i.e., any node $a \in H_i$ iff the distance from $a$ to some output node in $S$ is $i$. 
As a result, $H_{0} = Y = \{y_1, y_2, \ldots, y_n\}$.
Since $S'$ is an isomorphic-DRT-based structure, the nodes in $H_i$ have the same degree, which is referred to as $\rho_i$.
Our idea is to prove $|H_i| \geq n, \forall i \in [\zeta]$ such that $c(S') = \sum_{i =0}^{\zeta-1} |H_i|c_{\rho_i}\geq \sum_{i =0}^{\zeta-1} n c_{\rho_i}=\sum_{i=1}^{m-1} nw_i c_{i+1}$.

For any $a \in S'$, let $F(a) = (f_1, f_2, \ldots, f_{n})$, where for any $j \in [n]$, $f_j = 1$ if $x_j \in E(a, S')$ and $f_j = 0$ otherwise.
Meanwhile, for any $i \in [0,\zeta-1]$ and $H \subseteq H_i$, let $F(H) = \oplus_{a \in H} F(a)$, where $\oplus$ is the component-wise XOR operation.
If $H = \emptyset$, let $F(H) = (0, 0, \ldots, 0)$ ($n$ zeros in total).
Moreover, let $F(i) = \{F(H): H \subseteq H_i, |H| \text{~is even}\}, \forall i \in [0,\zeta-1]$.

On the one hand, for any $i \in [0, \zeta-2]$ and $ H = \{a_1, a_2, \ldots, a_k\} \subseteq H_i$ with even  $|H|$, we have $F(H)=F(a_1) \oplus \cdots \oplus  F(a_k)=(F(b_{1,1}) \oplus \cdots \oplus  F(b_{1,\rho_i})) \oplus \cdots \oplus  (F(b_{k,1}) \oplus \cdots \oplus  F(b_{k,\rho_i})) \in F(i+1)$, where $b_{k', j}$ is the $j$-th child of $a_{k'}$ in $E(a_{k'}, S')$ for $ k' \in [k], j\in[\rho_i]$.
That is, $F(0) \subseteq F(1) \subseteq \cdots \subseteq F(\zeta-1)$.
On the other hand, for any $H \subseteq H_0 = Y$ with even $|H|$, we have $F(H) = (f_1, f_2, \ldots, f_{n})$, where for any $j \in [n]$, $f_j = 1$ if $y_j \in H$ and $f_j = 0$ otherwise.
This leads to $|F(0)| = 2^{n-1}$.
As a result, we have $|F(i)| \geq 2^{n-1}, \forall i \in [0,\zeta-1]$, implying that $|H_i| \geq n$ as well as \eqref{eqn: T4}.

Let $S$ be the returned structure in Construction \ref{construction: latency optimal}.
Obviously, $S \in S^{\mathrm{isom}}(\mb w)$. 
Given \eqref{eqn: T4}, it suffices to prove $c(S) \leq \sum_{i=1}^{m-1}nw_ic_{i+1}$ to complete the proof of Theorem \ref{theorem: complexity of isomorphic DRT}.
Let $A_i$ denote the $i$-th level of $S$. 
By Construction \ref{construction:  latency optimal}, all DRTs $E(a, S)$, $a \in A_i$ are isomorphic  to each other and the leaves of each $E(a, S)$ are  labelled in a consecutive way, 
implying that $|A_i|=|\{E(a,S) : a\in A_i\}| \leq n$.
Then, $c(S) \leq \sum_{i=1}^{m-1}nw_ic_{i+1}$  follows.
\end{IEEEproof}

We remark that for $m = 2$, any isomorphic DRT is a perfect DRT that was considered in \cite{he22aclass}.
As a result, Theorem \ref{theorem: complexity of isomorphic DRT} can be thought as an extension of \cite[Theorem 6]{he22aclass} from $m = 2$ to $m \geq 2$.
Note that $\mb{w}$ can be any valid degree vector.
However, if $\mb w$ is the uniquely optimal solution to Problem 3,
$S$ has the lowest complexity among all the latency-optimal isomorphic-DRT-based structures.
On the other hand, if Problem 3 has multiple optimal solutions, we can enumerate them to find the complexity-optimal ones among latency-optimal isomorphic-DRT-based structures.

\section{Conclusion}\label{section: Conclusion}

In this paper, we have investigated the structures that can be used by a node in a message passing algorithm to compute outgoing messages from incoming messages.
Specifically, for the scenario where complexity has a higher priority than latency, we proposed a class of structures, called star-tree-based structures.
Within this framework, we derived algorithms to determine both the lowest achievable complexity and the corresponding lowest latency for these optimized structures.
Next, for the scenario where latency has a higher priority than complexity, we proposed a class of structures termed isomorphic-DRT-based structures.
We derived an algorithm to ascertain the lowest latency achievable with these structures and constructed latency-optimal isomorphic-DRT-based structures with the lowest complexity.
Notably, our findings extend the work presented in \cite{he22aclass} to accommodate multi-input scenarios.
Our future work is to investigate optimal/near-optimal structures that can achieve a good tradeoff between complexity and latency.

\ifCLASSOPTIONcaptionsoff
  \newpage
\fi

\bibliographystyle{IEEEtran}
\bibliography{myreference}

\end{sloppypar}
% that's all folks
\end{document}